\documentclass[a4paper,twoside]{article}

\usepackage{epsfig}
\usepackage{subcaption}
\usepackage{calc}
\usepackage{amssymb}
\usepackage{amstext}
\usepackage{amsmath}
\usepackage{amsthm}
\usepackage{multicol}
\usepackage{pslatex}
\usepackage{apalike}
\usepackage{algorithm2e}
\usepackage[bottom]{footmisc}

\usepackage{comment}
\usepackage{booktabs}
\usepackage{mathtools}
\usepackage{multirow,bigdelim}
\usepackage[capitalize]{cleveref}
\usepackage{tikz}
\usepackage{tikz-3dplot}
\usepackage{moresize}
\usepackage{footnote}
\usepackage{url}

\usepackage{SCITEPRESS}     % Please add other packages that you may need BEFORE the SCITEPRESS.sty package.

\newcommand{\norm}[1]{\left\lVert#1\right\rVert_2^2}
\newcommand{\euclideannorm}[1]{\left\lVert#1\right\rVert_2}

\DeclareMathOperator*{\argmax}{arg\,\max_{x}}

\definecolor{martincolor}{rgb}{0.8, 0.4, 0.1}

\definecolor{arnelacolor}{rgb}{0.0, 0.5, 0.0}
\newcommand{\arnela}[1]{\textcolor{arnelacolor}{[#1]}}

\begin{document}

\title{Teeth Localization and Lesion Segmentation in CBCT Images \\ using SpatialConfiguration-Net and U-Net}

%\author{\authorname{Anonymous Authors}}

\author{\authorname{
Arnela Hadzic\sup{1}%\orcidAuthor{0000-0000-0000-0000}
, 
Barbara Kirnbauer\sup{2}%\orcidAuthor{0000-0000-0000-0000}
, 
Darko \v{S}tern\sup{3}%\orcidAuthor{0000-0000-0000-0000} 
and 
Martin Urschler\sup{1}%\orcidAuthor{0000-0001-5792-3971}
}
\affiliation{\sup{1}Institute for Medical Informatics, Statistics and Documentation, Medical University of Graz, Graz, Austria}
\affiliation{\sup{2}Department of Dental Medicine and Oral Health, Medical University of Graz, Graz, Austria}
\affiliation{\sup{3}Institute of Computer Graphics and Vision, Graz University of Technology, Graz, Austria}
%\email{\{arnela.hadzic, barbara.kirnbauer, martin.urschler\}@medunigraz.at}
}

 
\abstract{The localization of teeth and segmentation of periapical lesions in cone-beam computed tomography (CBCT) images are crucial tasks for clinical diagnosis and treatment planning, which are often time-consuming and require a high level of expertise. However, automating these tasks is challenging due to variations in shape, size, and orientation of lesions, as well as similar topologies among teeth. Moreover, the small volumes occupied by lesions in CBCT images pose a class imbalance problem that needs to be addressed. In this study, we propose a deep learning-based method utilizing two convolutional neural networks: the SpatialConfiguration-Net (SCN) and a modified version of the U-Net. The SCN accurately predicts the coordinates of all teeth present in an image, enabling precise cropping of teeth volumes that are then fed into the U-Net which detects lesions via segmentation. To address class imbalance, we compare the performance of three reweighting loss functions. After evaluation on 144 CBCT images, our method achieves a 97.3\% accuracy for teeth localization, along with a promising sensitivity and specificity of 0.97 and 0.88, respectively, for subsequent lesion detection.}
%The abstract should summarize the contents of the paper and should contain at least 70 and at most 200 words. The text must be set to 9-point font size.

\onecolumn \maketitle \normalsize \setcounter{footnote}{0} \vfill

\section{\uppercase{Introduction}}
Cone-beam computed tomography (CBCT) is a highly effective medical imaging technique used to generate a 3D image of the oral and maxillofacial region, with various applications in dentistry~\cite{Khanagar21,Umer22}.
\begin{comment}
Cone-beam computed tomography (CBCT) is a highly effective medical imaging technique used to generate a 3D image of the oral and maxillofacial region. This technique has been widely applied in various fields of dentistry, including endodontics \cite{Umer22}, orthodontics \cite{Khanagar21}, and periodontics \cite{Chang20}, for clinical diagnosis and treatment planning. 
\end{comment}
However, the analysis and documentation of CBCT images typically require a significant amount of time and expertise from professionals.
While automated methods have been proposed to localize and segment anatomical structures in general medical images, this remains challenging for dental structures due to the presence of similar topologies and appearances among teeth, unclear boundaries, artifacts, or variations in shapes, appearance and size of lesions.
\begin{comment}
In recent years, various (semi-) automatic methods have been proposed to localize and segment anatomical structures in medical images, aiming to improve accuracy and reduce the time required for image interpretation. Machine learning techniques have also found applications in the field of dentistry. Nevertheless, automatic localization and segmentation of dental structures remain challenging due to the presence of similar topologies and appearances among teeth, unclear boundaries between them, metal artifacts, as well as variations in shapes, appearance, and sizes of lesions. 
\end{comment}
Moreover, in dental images lesions occupy a significantly smaller volume compared to the background and they only affect a small percentage of teeth in patients, resulting in class imbalance which poses another major challenge.
%\martin{Moreover, in dental images lesions occupy a significantly smaller volume compared to the background and they only affect a small percentage of teeth in patients, resulting in class imbalance which poses another major challenge.}

%\martin{Note: the following two paragraphs are very long, so the reader needs quite a while until they get to the contributions. An alternative way of presentation might be to have a short summary paragraph of related work here, followed by the contributions, and then a more detailed related work sub-section or even chapter after that. But I also think that for this paper and venue, it might be fine as it is. So your choice, how you would like to have it.}

% LOCALIZATION
Early approaches for localizing anatomical structures in medical images were based on statistical models of shape and appearance \cite{Cootes94}, which were later improved by incorporating random forest-based machine learning models~\cite{Donner13,Unterpirker2015-zp,Urschler2018-eq}.
\begin{comment}
However, the main limitation of these models was their linearity in the shape modeling. Later developments in machine learning focused on improving existing methods for landmark localization by incorporating random forests \cite{Donner13,Unterpirker2015-zp}.
\end{comment}
\begin{comment}
For example, \cite{Criminisi13} use an ensemble of regression trees to estimate the position of bounding boxes around organs in 3D CT images. On the other hand, \cite{Donner13} combine Random Forest classification with a Markov Random Field for localizing anatomical landmarks. Similarly, \cite{Lindner14} combine Random Forest regression with a statistical shape model for landmark detection.  Both methods showed an improved accuracy of a single classifier/regressor by incorporating global shape information. For the detection of dental landmarks in 3D CBCT images, \cite{Cheng11} use a Random Forest combined with sampled context features to constrain the region in which the landmarks should be searched for. Differently, \cite{Unterpirker2015-zp} propose using Random Regression Forests to detect third molars in 3D magnetic resonance images.
\end{comment}
Recently, deep learning-based methods have outperformed traditional machine learning localization approaches in terms of accuracy and efficiency. \cite{Zhang17} proposed two deep convolutional neural networks (CNNs) for detecting anatomical landmarks in brain MR volumes, where the first network captures inherent relationships between local image patches and target landmarks, while the second network predicts landmark coordinates directly from the input image. \cite{Jain13}  found that regressing heatmaps rather than coordinates improves the overall performance of landmark detection and also simplifies the analysis of network's predictions. Building on this idea, \cite{Payer19} proposed the SpatialConfiguration-Net, which combines local appearance responses with the spatial configuration of landmarks in an end-to-end manner and achieved state-of-the-art performance on a variety of medical datasets. 
\begin{comment}
The SpatialConfiguration-Net has recently been used in a cascade manner to localize bone anatomical landmarks in spinal CT images \cite{Zhao22}.
\end{comment}
\begin{comment}
, but was also successfully evaluated for facial and human pose landmark localization in natural images \cite{Payer2019Evaluating}, thus demonstrating its versatility. 
\end{comment}
%\martin{, but also was successfully evaluated for facial and human pose landmark localization in natural images cite Payer2019 IVCNZ DOI: 10.1109/IVCNZ48456.2019.8961000, thus demonstrating its versatility}. 
Regarding teeth detection in CBCT images, several studies that use CNNs have been published. 
\begin{comment}
\cite{Miki17} employed a fully convolutional network based on the AlexNet architecture %\martin{Achtung: ich glaube nicht dass AlexNet fully convolutional ist, es ist ein netz für klassifikation. FCNs kamen erst danach auf...} 
to locate teeth regions. % accuracy: 77.4\%
\end{comment}
\begin{comment}
\cite{Zakirov18} trained a V-Net-based fully convolutional network to detect bounding boxes of present teeth, formulating the problem as a 33-class semantic segmentation. 
\end{comment}
% accuracy: 96.3%, classification instead of object detection (exact location detection)
\cite{Chung20} adopted a faster R-CNN framework to localize individual tooth regions inside volume of interest regions, which were previously extracted and realigned using a 2D pose regression CNN. % additional network used for VOI
More recently, \cite{Du22} employed a classification network to extract tooth regions and trained a YOLOv3 network to detect teeth bounding boxes within these regions.

% SEGMENTATION
Early approaches for segmenting structures in medical images often relied on active contours or statistical shape models \cite{Heimann09,Gan17}. 
\begin{comment}
\cite{Heimann09,Gollmer12,Gan17}
\end{comment}
Their limitation of requiring predefined handcrafted features was recently overcome by deep learning-based methods.
\begin{comment}
However, these methods require predefined handcrafted features, which significantly affects the accuracy of the segmentation results. In recent years, deep learning-based methods have become the dominant approach in various computer vision tasks, including medical image segmentation \cite{Wang22}.
\end{comment}
To date, several research studies have been proposed for the automated detection of periapical lesions in dental images using deep learning. Most of these studies focus on periapical or panoramic radiographs \cite{Ekert19,Endres20,Krois21,Pauwels21}, which exhibit lower accuracy compared to using CBCT scans \cite{Antony20}, potentially leading to missed or hidden lesions. To the best of our knowledge, only a limited number of research studies have addressed the automated detection or segmentation of periapical lesions in CBCT images. \cite{Lee20} employed a CNN architecture based on the GoogLeNet Inception v3 model and trained it on cropped 2D slices from CBCT images. They achieved a sensitivity value of 0.94 for periapical lesion detection.
\cite{Setzer20} used a U-Net-based architecture to segment periapical lesions in limited field-of-view CBCT data. Their model achieved a sensitivity of 0.93 and a specificity of 0.88, with an average Dice score of 0.52 for all positive examples and 0.67 for true positive examples. However, their training and testing involved 2D scans from only 20 CBCT volumes with 61 roots. In another study, \cite{Zheng20} trained an anatomically constrained Dense U-Net using 2D slices from 20 CBCT images. They incorporated oral-anatomical knowledge that periapical lesions are located near the roots of teeth and achieved a sensitivity value of 0.84. \cite{Orhan20} employed two separate U-Net-like CNN architectures for teeth localization and periapical lesion segmentation in CBCT images. The first network localized each tooth, and the second network used the extracted teeth with their context to detect lesions. However, the authors did not provide details about the architecture or the training/testing procedure of their networks. They reported detecting 142 out of 153 lesions correctly, resulting in a sensitivity of 92.8\% for lesion detection, with only one misidentified tooth. They did not provide an evaluation of negative examples, nor did they clarify whether 3D volumes were used throughout their method.

Although deep learning techniques have demonstrated potential in automating the detection of periapical lesions in CBCT images, 
\begin{comment}
more extensive research and evaluation with larger datasets are necessary to enhance their accuracy and reliability. 
\end{comment}
current methods often rely on training procedures based solely on 2D slices, which may result in the loss of valuable information. To address this concern, it is important to incorporate 3D volumes into the training process to effectively utilize all available data. Additionally, the issue of class imbalance should be taken into account, since lesions occupy only small volumes in images, and the majority of teeth are lesion-free.

%\paragraph{Contribution.}
In this work, we have developed a fully automated deep learning method for the detection of teeth and periapical lesions in 3D CBCT images in a multi-step process. First, we use the 3D SpatialConfiguration-Net to perform the teeth localization, i.e., to generate a 3D coordinate of each tooth in an image. Then, we automatically crop each tooth in the image based on the generated coordinates. 
\begin{comment}
This cropping is done in such a way that the center coordinate of the cropped volume corresponds to the generated coordinate of the respective tooth.
\end{comment}
Finally, we train a 3D U-Net using the previously cropped volumes to segment periapical lesions. To address the commonly encountered class imbalance problem in medical datasets, we use and compare three different reweighting loss functions during both the training and testing procedures. Utilization of state-of-the-art (SOTA) 3D network architectures for respective tasks as well as adaptation to address class imbalance problems contribute to the reliability and high accuracy of our method. The objective of this study is to provide a comprehensive description of the techniques used to obtain the results that we recently published in a clinical journal \cite{Kirnbauer22}, and to shed light on the significant issues of class imbalance and preservation of 3D volumetric information.

%\martin{preservation of volumetric information}.

%\martin{die einführung des Endodontics paper finde ich gut gelungen! auch das class imbalance problem ist ausreichend motiviert an dieser stelle. Aber der issue, dass die meisten Studien 2D Netze benutzt haben, ist noch nicht stark in der related work unterstrichen. dein summary paragraph nach related work sollte nochmal betonen, dass die bisher reporteden arbeiten ausschliesslich auf 2D slices trainiert wurden. dieser aspekt gehoert dann später auch noch diskutiert (siehe kommentare später).}

\section{\uppercase{Method}}
\subsection{Data}
Our method was trained and tested on a dataset that consists of 144 3D CBCT images provided by the University Clinic for Dentistry and Oral Health Graz. Ethical approval was granted by the Medical University of Graz under review board number 33-048 ex 20/21. Out of the 144 images, 16 images visualize both jaws, while the remaining 128 images visualize either the upper or lower jaw. 
The images visualize 2128 teeth, and in most images at least one periapical lesion was found. 
In total, approximately 10\% of the teeth in the dataset were affected by a periapical lesion, thus leading to class imbalance.

\begin{comment}
\subsection{Ground Truth Data}
\end{comment}
To obtain ground truth data, we first perform a manual localization of each tooth location, thus creating a set of 32 coordinates for each image. If a tooth is missing, we annotate it with the coordinate $(-1, -1, -1)$.
\begin{comment}
In order to obtain ground truth data, we perform a manual localization of each tooth and a semi-automatic binary segmentation of each periapical lesion in our dataset. To create the ground truth landmarks of the teeth locations in an image, we used the Medical Imaging Interaction Toolkit (MITK) \footnote{\url{https://www.mitk.org/wiki/The_Medical_Imaging_Interaction_Toolkit_(MITK)} (accessed 27.09.2023).}. For each image, we create a set of 32 coordinates. If a tooth is missing, we annotate it with the coordinate $(-1, -1, -1)$. Otherwise, the ground truth coordinate corresponds to the actual tooth location in the image.
\end{comment}
Second, for the ground truth segmentation of periapical lesions in an image, we use the semi-automatic Total Variation (TV) framework proposed in \cite{Urschler14}. 
\begin{comment}
Based on the user's initial contours in the image regions affected by periapical lesions, the TV software segments lesions automatically iteratively by minimizing the convex energy functional 
\begin{equation}
    \label{equation:TV_energy_functional}
    \min_{u} 
    { 
        \bigg \{ \int_{\Omega}g(x) |\nabla u(x)|\mathop{dx} + \lambda \int_{\Omega}u(x)f(x)\mathop{dx} \bigg \}
    },
\end{equation}
where
\begin{equation}
    \label{equation:gradients_formula}
    g(x)=\mathrm{exp}^{-\alpha |\nabla I(x)|^\beta}.
\end{equation}

In the above equations, $\Omega$ represents the image domain, $g$ an edge-detection function, $x\in\Omega$ image voxels, $f$ a user-provided input, and $u$ a generated solution, i.e., a generated segmentation map. The first term of the energy functional corresponds to the weighted TV of $u$, while the second term integrates constraints into the energy functional. The expression $f(x)$ can be either provided by the user or derived from a prior classification into foreground ($f=1$) and background ($f=0$) pixels. The parameter $\lambda$ is employed to interpret the information contained in $f$. The function $g(x)$ describes the image gradients, where $\alpha$ and $\beta$ are tunable parameters used to map intensity image gradients to the range $[0, 1]$. We set the parameters $\lambda=50$, $\alpha=15$, and $\beta=0.55$. 
\end{comment}
Each lesion segmentation was then verified by an experienced dentist using the ITK-Snap software \cite{py06nimg} and manually adjusted if necessary.

\begin{figure*}[t]
\centering
\scalebox{1.6}{
   {\input{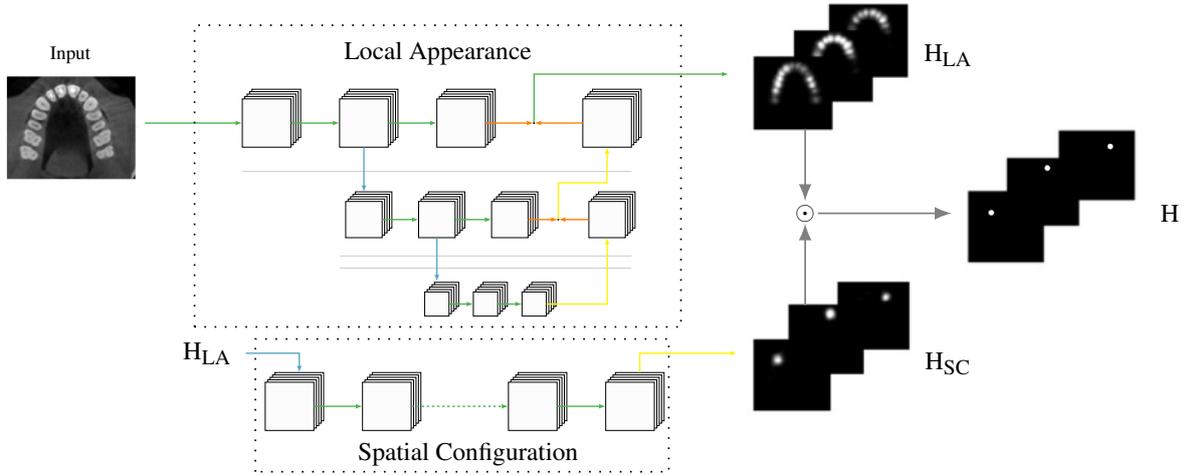}}}
  \caption{Teeth localization using SpatialConfiguration-Net. The local appearance component generates a set of local appearance heatmaps denoted as $\mathrm{H_{LA}}$, while the spatial configuration component produces a set of spatial configuration heatmaps $\mathrm{H_{SC}}$. The final heatmap images $\mathrm{H}$ for teeth landmarks are obtained by performing voxel-wise multiplication of $\mathrm{H_{LA}}$ and $\mathrm{H_{SC}}$. The arrows represent different operations: green arrow -- convolution, blue arrow -- downsampling, yellow arrow -- upsampling, and orange arrow -- voxel-wise addition.
  %\martin{Note: auf meinem paper ausdruck sieht man die landmark locations ganz rechts sehr schwer. vielleicht kann man das noch färbig machen? Man koennte weiters einen hinweis bei der caption hinzufügen, dass die beste visuelle qualität dieses bildes im pdf sichtbar ist (best viewed in the pdf version).}
  }
  \label{fig:scn_architecture}
 \end{figure*}

\subsection{CNN Architecture}
\begin{figure*}[t]
\begin{centering}
\scalebox{0.36}{
  \centering
   {\input{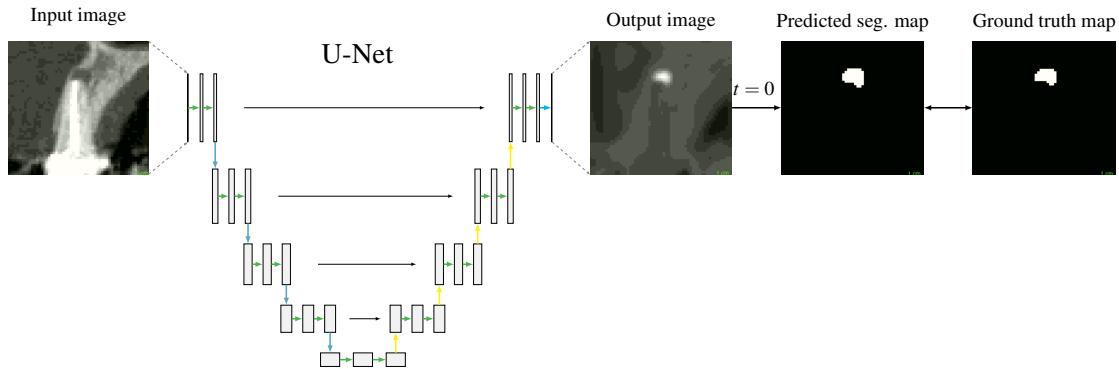}}}
  \caption{Periapical lesion segmentation using a modified U-Net architecture. A cropped image of a tooth along with its corresponding ground truth map is used as input to the U-Net. The network is trained to output an image of intensity values, where each value lies in the range $(-\infty, +\infty)$. A threshold $t$ is applied to the predicted intensity image in order to generate a final binary segmentation map.
  %\martin{in meinem ausdruck von der letzten paper version war dieses bild auf der 6. seite. schau bitte, dass dieses bild im final paper auf der 5. seite (oder jedenfalls auf der nächsten seite von Fig. 1) zu liegen kommt.}
  }
  \label{fig:unet_architecture}
\end{centering}
\end{figure*}

The developed method consists of two networks (see \cref{fig:scn_architecture,fig:unet_architecture}): the SpatialConfiguration-Net \cite{Payer19}, and a modified version of the U-Net \cite{Ronneberger15}. The method includes an additional cropping step in between, a design inspired by \cite{Payer20}.
\begin{comment}
a design which was inspired by \cite{Payer20} who localized and segmented spine vertebrae in CT images.
\end{comment}
% \martin{a design which was inspired by Payer et al. 2020 (das VISAPP paper spine localization and segmentation DOI: 10.5220/0008975201240133) who localized and segmented spine vertebrae in CT images.}. 
\begin{comment}
The networks are schematically visualized in \cref{fig:scn_architecture} and \cref{fig:unet_architecture}.
\end{comment}

\subsubsection{Teeth Localization} 
As the first step towards the detection of periapical lesions in 3D CBCT images, we perform the teeth localization using the SpatialConfiguration-Net (SCN). SCN is a fully convolutional neural network that consists of two main components: \textit{local appearance} (LA) and \textit{spatial configuration} (SC). The LA component generates locally accurate predictions, but they may be ambiguous. To solve this issue, the SC component incorporates the spatial relationship between landmarks into the network. 

The network is trained to produce heatmap images of landmarks, where each heatmap image represents the probability of a specific landmark being present at a particular location in the image. The predicted coordinate $x'_i$ of a landmark $L_i$, where $i\in\{1, \dots, 32\}$, is defined as the coordinate where the predicted heatmap $h_i(x;w,b)$ has its highest value.
\begin{comment}
, i.e.,
\begin{equation}
    \label{equation:argmax}
    x'_i = \argmax {h_i(x; w, b)} .
\end{equation}
\end{comment}

For each image in our dataset, we create a target heatmap image $g$ by merging Gaussian heatmaps of its ground truth landmarks. A Gaussian heatmap $g_i(x;\sigma_i)$ of a ground truth landmark $L_i$ is defined by
\begin{equation}
    \label{equation:gaussianfunction}
    g_i(x; \sigma_i) = \exp\left(- \frac{\norm{x - \tilde{x}_i}} {2 \sigma_i^2} \right) ,
\end{equation}
where $x$ are image coordinates and $\tilde{x}_i$ is the ground truth coordinate of the landmark $L_i$. The heatmap peak widths are determined by the standard deviation $\sigma$ and depend on the distance between the image coordinates and the ground truth coordinate. Higher values are assigned to voxels that are closer to the ground truth coordinate $\tilde{x}_i$, while the values of voxels further away from $\tilde{x}_i$ decrease gradually.

To minimize differences between the predicted $h_i(x; w, b)$ and the target heatmaps $g_i(x; \sigma_i)$ for each landmark $L_i$, we minimize the objective function
\begin{equation}
    \label{equation:lossfunction}
    \min_{w, b, \sigma} \sum_{i=0}^{N-1} \sum_{x} \norm{h_i(x; w, b) - g_i(x; \sigma_i)}\cdot M(x) + T, 
\end{equation}
where
\begin{equation}
    \label{equation:lossfunction_additionalterm}
    T =\alpha \norm{\sigma} + \lambda \norm{w}.
\end{equation}
We calculate the distance between the predicted heatmaps and the target heatmaps using the $L_2$ measure, which is multiplied with a binary mask $M(x)$. When the ground truth annotation at location $x$ is annotated as missing, the value of $x$ in $M(x)$ is set to zero. This way, the network ignores the predictions for missing teeth. The heatmap peak widths $\sigma = (\sigma_0, \sigma_1, \dots, \sigma_{N-1})^T $, the network weights $w$, and the bias $b$ are learnable parameters of the network. The factor $\alpha$ determines how strong the heatmap peak widths $\sigma$ are being penalized, while $\lambda$ determines the impact of the $L_2$ norm of the weights $w$. 
\begin{comment}
In order to minimize the objective function in \labelcref{equation:lossfunction}, the first term should have a large $\sigma$ value, while the second term should have the smallest possible $\sigma$ value. Since these terms oppose each other, the network has to learn the optimal $\sigma$ value \cite{Payer19}. 
\end{comment}

The LA component consists of four levels, where each level includes three convolutional layers and one average pooling layer, except for the last level where downsampling is not performed. The SC component consists of one level with four 7x7x7 convolutional layers, three of which have 64 outputs and one has 32 outputs. The inputs to the first convolutional layer are the local appearance heatmaps $\mathrm{H_{LA}}$ generated by the LA component, downsampled by a factor of 4. To generate the set of spatial configuration heatmaps $\mathrm{H_{SC}}$, the 32 outputs of the last convolutional layer are upsampled to the input resolution using tricubic interpolation. All convolutional layers, except the ones generating $\mathrm{H_{LA}}$ and $\mathrm{H_{SC}}$, have a LeakyReLU activation function with a negative slope of 0.1. The weights are initialized using the He initializer \cite{He2015-hz}. The layer generating $\mathrm{H_{LA}}$ has a linear activation function, while the layer generating $\mathrm{H_{SC}}$ has a TanH activation function. Both layers initialize the biases with 0 and the weights using a Gaussian distribution with a standard deviation of 0.001. 
\begin{comment}
The biases in all convolutional layers are initialized with 0. 
To prevent overfitting, 
\end{comment}
A dropout rate of 0.3 is applied after the first convolution layer in each level.

As shown in \cref{fig:scn_architecture}, the LA and SC components generate separate heatmap images, which are then multiplied voxel-wise to generate the final output of the SCN. Using the coordinates predicted by the SCN, all teeth present in the images are cropped to the size of $[64, 64, 64]$ and fed into the U-Net for lesion segmentation.

\subsubsection{Lesion Segmentation}

For the segmentation of periapical lesions, we use a modified version of the U-Net. Our adaptation consists of 5 levels, where each convolutional layer has a kernel size of $[3, 3, 3]$ and 16 filter outputs. In the contracting path, we use convolutional operations followed by downsampling through average pooling. In the expansive path, each upsampling layer performs trilinear interpolation followed by two convolutional operations. After each convolutional layer we apply a dropout of 0.3, and use 'same' padding to maintain the same input and output size. All layers have a ReLU activation function except the last one, where no activation function is used in order to obtain logits instead of probabilities as the network's output. Same as for the SCN, the weights in the last layer are initialized with the He initializer. In all other layers, the initial weights are sampled from a truncated normal distribution with a standard deviation of 0.001. The last layer consists of a single output, generating an image with predicted intensity values. The network's output is an image of size $[64, 64, 64]$, with voxel intensities in the range $(-\infty, +\infty)$. Finally, a threshold of 0 was applied to the output in order to produce the predicted binary segmentation map, where all voxel values below the threshold are considered background, while non-negative values represent a detected lesion.

To address the class imbalance problem, we compare three objective functions, Focal Loss FL \cite{Lin17}, Focal Tversky Loss FTL \cite{Abraham19}, and Combo Loss CL \cite{Taghanaki19} based on Dice Similarity Coefficient (DSC):
\begin{equation}
    \mathrm{FL}(p_t) = - \alpha_t(1-p_t)^\gamma \log(p_t),
\end{equation}
\begin{equation}
    \label{equation:obj_function_focal_tversky_loss}
    \mathrm{FTL}(P, G) = (1 - \mathrm{TI}(P, G))^{\frac{1}{\gamma}},
\end{equation} 
and
\begin{equation}
    \mathrm{CL}(y, p) = \delta \cdot L_\mathrm{BBCE}(y,p) - (1-\delta) \cdot \mathrm{DSC}(y,p).
\end{equation} 
In the above equations, Tversky Index (TI) and Balanced Binary Cross-Entropy (BBCE) are defined by
\begin{equation}
    \label{equation:obj_function_tversky_index}
    \mathrm{TI}(P, G) = \frac{|PG| + \epsilon}{|PG|+\beta |P\backslash G|+(1-\beta) |G\backslash P| + \epsilon}
\end{equation} 
and
\begin{equation}
    L_{\mathrm{BBCE}}(y, p) = -\alpha \cdot y \log(p) - (1-\alpha) \cdot (1-y) \log(1-p).
\end{equation}
%DSC is the Dice Similarity Coefficient.
The parameters $\alpha$ and $\beta$ in the above equations are used to weight positive and negative examples. $\alpha$ can be set as the inverse class frequency or considered as a hyperparameter, while values of  $\beta$ larger than 0.5 give more significance to false negative examples. To enhance the focus on misclassified predictions, $\gamma$ should be larger than 0 in the FL function and larger than 1 in the FTL function. The parameter $\delta$ regulates the contribution of the BBCE to the CL function.

%\section{\uppercase{Evaluation}}
\subsection{Data Augmentation}
To prepare images for training the SCN for localization, we resize all images from the original size of $[501, 501, 501]$ to a new size of $[64, 64, 32]$, while maintaining a fixed aspect ratio. As we do not need highly precise locations of teeth for the lesion segmentation task, the SCN can be trained on downsampled images. Image intensities are then scaled to the range $[-1, 1]$. Additionally, images undergo translation, rotation, and scaling using random factors sampled from uniform distributions ($[-10, 10]$ for translation, $[-0.1, 0.1]$ for rotation, $[0.9, 1.1]$ for scaling). 
\begin{comment}
The randomly generated translation factors lie within the range $[-10, 10]$, rotation factors within $[-0.1, 0.1]$, and scaling factors within $[0.9, 1.1]$. 
\end{comment}

Before training our modified U-Net for the lesion segmentation task, original high-resolution images are cropped along with their corresponding ground truth segmentation maps for each tooth. This process generates 32 cropped images from a single original image, where the center of each cropped image corresponds to the center coordinate of a particular tooth. Each cropped image has dimensions of $[64, 64, 64]$ and a spacing of $[0.4, 0.4, 0.4]$. Since periapical osteolytic lesions are always located in the area of the root tips, we translate cropped images by fixed factors to ensure visibility of the entire tooth root within a cropped image. The translation factors of $[0, 1, -9]$ are applied to the cropped images with a tooth in the upper jaw. The cropped images with a tooth in the lower jaw are translated by factors $[0, 3, -9]$ and rotated by 180 degrees. Moreover, all cropped images undergo random translation within the range of $[-1, -1]$, random rotation, and random scaling withing the range of $[-0.2, 0.2]$. As a post-processing step, we normalize cropped images to the range $[0, 1]$ and perform shift scale intensity transformation using a random factor of 0.6. We perform label smoothing by applying a Gaussian kernel with a standard deviation of 0.1 to the ground truth segmentation maps. 
\begin{comment}
Data augmentation was done using SimpleITK.
\end{comment}

\subsection{Implementation Details}
An Intel(R) CPU 920 with an NVIDIA GeForce GTX TITAN X was used for the training and testing of the model. The CNNs were running on Ubuntu 20.04 operating system with Python 3.7 and TensorFlow 1.15.0. To evaluate the performance of our model, we used 4-fold cross validation (cv). Total training time of the SCN for one cv fold took about 20 hours, while one fold for training the U-Net took about 17 hours. The teeth localization network took around 15 seconds for inference on a CBCT volume, while the segmentation network, including cropping, took approximately 2 to 3 minutes.

To minimize the loss function during training of SCN, we use the Nesterov Accelerated Gradient with a learning rate of $10e-7$ and momentum value of 0.99. We set the number of iterations to 15,000 as we did not observe any substantial improvement after that. Remaining hyper-parameters batch size, weight decay, and sigma regularization term are determined empirically and set to 1, $5e-5$, and 100, respectively.

The U-Net is trained for 113,000 iterations using a batch size of 8. As mentioned earlier, we address the class imbalance issue by utilizing Focal Loss, Focal Tversky Loss, and Combo Loss. We set the parameters of these loss functions as follows:
Within our dataset, positive voxels in each cropped image occupy less than 10\% of the volume. Therefore, we set the weighting factors $\alpha$ and $\beta$ to 0.9 in the corresponding loss functions. The parameter $\gamma$ controls the down-weighting of easy examples. We set $\gamma$ to 2 in the Focal Loss and the Focal Tversky Loss function. In the case of the Combo Loss function, we set the parameter $\delta$ to 0.5, which determines the contribution of the Balanced Binary Cross-Entropy. Furthermore, in order to minimize the loss functions, we employ the Adam optimizer with a learning rate of $1e-4$.

Both networks are trained and tested using the 4-fold cross validation technique. For teeth localization, our dataset of 144 3D CBCT images is divided into four cv folds, each containing 36 images. For lesion segmentation, we split a total of 2128 cropped images into four cv folds. Since only 206 teeth in our dataset are affected by a lesion, we distribute them uniformly over all cv folds, resulting in 10\% of teeth with lesions and 90\% teeth without lesions per fold. 

\subsection{Metrics}
To evaluate the teeth localization performance, we use the point-to-point error (PE) and accuracy metrics. For a landmark $i$, the point-to-point error $\text{PE}_i$ represents the Euclidean distance between the ground truth landmark $\tilde{x}_i$ and the predicted landmark $x'_i$.
\begin{comment}
for an image $i$, defined by the formula
\begin{equation}
    \label{equation:PE}
    \text{PE}_i^{(j)} = \euclideannorm{\tilde{x}_i^{(j)} - x_i^{'(j)}}.
\end{equation}
\end{comment}
Accuracy is defined as the percentage of correctly identified landmarks over all predicted landmarks (i.e., average precision). 
\begin{comment}
, i.e.,
\begin{equation}
    \label{equation:accuracy}
    \text{Accuracy} = \frac{\#\text{ID}_\text{correct}}{\#\text{all\_landmarks}} ,
\end{equation}
\end{comment}
A predicted landmark $x'_i$ is identified correctly if the closest ground truth landmark is the correct one and the Euclidean distance to the ground truth landmark is less than a specified radius $r$.

To evaluate the lesion detection performance, 
%we formulate the lesion detection problem as a classification task, where we report for each tooth whether it is affected by a lesion or not. For this purpose, 
we use the metrics sensitivity $(\text{TP} / (\text{TP} + \text{FN}))$ and specificity $(\text{TN} / (\text{TN} + \text{FP}))$. Each predicted segmentation is evaluated based on the Dice score, an overlap measure between the ground truth segmentation map $X$ and the predicted segmentation map $Y$, defined by
\begin{equation}
    \label{equation:dice}
    \text{DSC} = 2\cdot\frac{\lvert X \cap Y \rvert}{\lvert X \rvert + \lvert Y \rvert}.
\end{equation}
\begin{comment}
evaluation metrics, defined by
\begin{equation}
    \label{equation:sensitivity}
    \text{Sensitivity} = \mathrm{\frac{TP}{TP+FN}}
\end{equation}
and
\begin{equation}
    \label{equation:specificity}
    \text{Specificity} = \mathrm{\frac{TN}{TN+FP}}.
\end{equation}
\end{comment}
Sensitivity represents the network's ability to correctly detect teeth with lesions, whereas specificity represents its ability to correctly detect teeth without lesions. A predicted segmentation map for a tooth with a lesion is classified as a True Positive (TP) if the Dice score between the ground truth and  predicted segmentation map is larger than 0. Otherwise, the predicted segmentation map is classified as a False Negative (FN). A predicted segmentation map for a tooth without a lesion is classified as a True Negative (TN) if all voxels in the predicted segmentation map are 0. Otherwise, the predicted segmentation map is classified as a False Positive (FP) prediction. 

%The overlap between the ground truth segmentation map $X$ and the predicted segmentation map $Y$ is evaluated using the Dice score

\subsection{Delaunay Triangulation}
For the training of the U-Net, each original image is cropped to the size of $[64, 64, 64]$ for each tooth, where the center coordinate of a cropped image corresponds to the center coordinate of the particular tooth in the original image. Since teeth can have different shapes, sizes, and orientations, the resulting cropped images may contain multiple teeth, which complicates automated tooth-based evaluation. To address this issue, we use the definitions of the convex hull and the Delaunay triangulation.

First, we annotate a cuboid around each tooth in the original image using Planmeca Romexis\textsuperscript{\textregistered} software. This annotation ensures that each cuboid encloses only the area of a single tooth. The software then automatically generates a set of 3D coordinates representing the cuboid. We use these coordinates as input for the 'Convex-Hull' and 'Delaunay' functions from the SciPy 1.6.2 library.
\begin{comment}
, which compute the convex hull and the Delaunay triangulation of the provided coordinates. 
\end{comment}
By using the ’Delaunay.find simplex’ function, we obtain the indices of all the simplices that contain all cuboid voxels. Using these indices, we are able to generate a segmentation map for an annotated cuboid. Finally, during the evaluation of a specific tooth, we only consider the voxels that belong to its corresponding annotated cuboid.

\section{\uppercase{Results and Discussion}}
The results of the teeth localization and lesion segmentation tasks are shown in \cref{tab:teeth_localization_results} and \cref{tab:lesion_detection_results}, respectively. Moreover, we provide a comparison of the results with SOTA methods for periapical lesion detection in \cref{tab:sota_comparison}.

%The SCN was evaluated using the point-to-point error and accuracy metrics for the radii of 2, 3, 4, and 6 mm. The 4-fold cross validation results for teeth localization are shown in  \cref{tab:teeth_localization_results}.
\begin{comment}
The SCN achieved an average accuracy of 97.3\% by utilizing a radius $r$ of 6 mm to define the correctly identified landmarks. In other words, for 97.3\% of the predicted teeth, the closest ground truth landmark was the correct one and the distance between the predicted and ground truth landmarks was less than 6 mm. Decreasing the radius to 4, 3, and 2 mm resulted in average accuracy values of 94.7\%, 89.1\%, and 72.6\%, respectively. 
An overview of the teeth localization results is shown in \cref{tab:teeth_localization_results}.
\end{comment}

\begin{table}[h]
    \vspace{0.3cm}
    \caption{Teeth localization results for 4-fold cross validation, showing point-to-point error with standard deviation, as well as accuracy of the model for different radii $r$.}\label{tab:teeth_localization_results} \centering
    \begin{tabular}{c|lc|c}
        \hline
        \multirow{2}{*}{Radius $r$} & & PE in mm      & \multirow{2}{*}{Accuracy (\%)} \\
                                    & & Mean $\pm$ SD &  \\
        \hline
        \hline
        2 mm & \rdelim\}{4}{3mm}[] & & 72.6  \\
        3 mm & & 1.74 $\pm$ 1.44 & 89.1 \\
        4 mm & & & 94.7  \\
        6 mm & &  & \textbf{97.3}  \\
       \hline
    \end{tabular}
\end{table}

%The modified U-Net was trained with three different loss functions and evaluated using the Sensitivity, Specificity, and Dice Score metrics. 
\begin{comment}
When the modified U-Net trained with Focal Loss was used for lesion segmentation, the sensitivity and specificity metrics had mean values of $0.97\pm 0.03$ and $0.88 \pm 0.04$ respectively. 
This means that 97.1\% of the lesions were correctly detected as True Positives, while 2.9\% of lesions were missed as False Negatives. Moreover, 88\% of teeth without lesions were correctly detected as True Negatives, whereas 12\% of teeth without lesions were incorrectly predicted as False Positives. 
The corresponding average Dice score value was \arnela{$0.67 \pm 0.03$}. The U-Net trained with Focal Tversky Loss achieved a sensitivity value of \arnela{$0.92 \pm 0.05$} and a specificity value of \arnela{$0.89 \pm 0.08$}, with a Dice score value of \arnela{$0.70 \pm 0.04$}. The same Dice score value was obtained by the U-Net trained with the Combo Loss, but with a lower sensitivity value of \arnela{$0.85 \pm 0.04$} and a higher specificity value of \arnela{$0.94 \pm 0.02$}. 
\cref{tab:lesion_detection_results} provides an overview of the lesion detection and segmentation results based on the coordinates predicted by the SCN.
\end{comment}

\begin{table}[h]
    \vspace{0.1cm}
    \caption{Lesion detection and segmentation results for 4-fold cross validation, shown as sensitivity, specificity, and Dice score obtained using different loss functions.}\label{tab:lesion_detection_results} \centering
    \begin{tabular}{c|c|c|c}
        \hline
        Loss & Sensitivity & Specificity & DICE \\
        \hline
        \hline
        FL & \textbf{0.97} $\pm$ 0.03 & 0.88 $\pm$ 0.04 & 0.67 $\pm$ 0.03 \\
        FTL & 0.92 $\pm$ 0.05 & 0.89 $\pm$ 0.08 & \textbf{0.70} $\pm$ 0.04 \\
        CL & 0.85 $\pm$ 0.04 & \textbf{0.94} $\pm$ 0.02 & \textbf{0.70} $\pm$ 0.04 \\
        \hline
    \end{tabular}
\end{table}

\begin{comment}
In \cref{tab:sota_comparison}, we present the comparison results of our method with state-of-the-art techniques for periapical lesion detection. 
\end{comment}

\begin{table}[h]
    \vspace{0.1cm}
    \caption{Comparison with SOTA methods.}\label{tab:sota_comparison} \centering
    \begin{tabular}{c|c|c}
        \hline
        Method & Sensitivity & Specificity \\
        \hline
        \hline
        \cite{Lee20} & 0.94 & -  \\
        \cite{Setzer20} & 0.93 & \textbf{0.88} \\
        \cite{Zheng20} & 0.84 & - \\
        \cite{Orhan20} & 0.93 & - \\
        \textbf{Ours} & \textbf{0.97} & \textbf{0.88} \\
        \hline
    \end{tabular}
\end{table}

%\martin{hier auch erwähnen, dass diese modelle auf 2D slices arbeiten, im gegensatz zu unseren, und dass deren (wie ich vermute) trainings und test datensets kleiner sind als unserer}

%\martin{Note: Mach nochmal für dich den check: Die Intro soll den weg bereiten, warum deine contributions (class imbalance, 3D processing vs. 2D processing) relevant sind. dies kann über argumente und/oder sich beziehen auf related work erfolgen. die contributions section sollte dann deine contributions aus obiger motivation definieren. die methode sollte erklären, auf welche art und weise diese contributions umgesetzt wurden. die experimente machen klar, wie sich diese umgesetzten modifikationen auf performance auswirken, bzw. im vergleich zu state of the art dastehehn. die discussion und conclusion erläutern dies dann nochmal im detail, sodass der leser zufrieden sein kann, dass diese contributions meaningful sind. Ich glaube du hast das im grossen und ganzen erfüllt, aber es macht sinn wenn du dir das nochmal verdeutlichst und checkst, ob dieser rote faden durchgeht, und für den leser verständlich ist. }

\begin{comment}
\section{\uppercase{Discussion and Conclusion}}
\end{comment}

We evaluated the performance of the SCN on 4 folds and achieved an average accuracy of 97.3\% in teeth detection, with a mean point-to-point error of 1.74 $\pm$ 1.44 mm. In other words, for 97.3\% of the predicted teeth, the closest ground truth landmark was the correct one and the distance between the predicted and ground truth landmarks was less than 6 mm. Decreasing the radius to 4, 3, and 2 mm resulted in average accuracy values of 94.7\%, 89.1\%, and 72.6\%, respectively.
By observing individual predictions, we could notice that the most incorrect predictions occurred in images affected by artifacts or images with misaligned teeth. This can be attributed to these cases being rare in our dataset and thus deviating from the learned data distribution. In future work, this could be addressed by combining generative models with CNNs to incorporate global shape/landmark configurations into the training process. This way, the detection of out-of-distribution data could be improved.

The U-Net was trained using the Focal Loss (FL), Focal Tversky Loss (FTL), and Combo Loss (CL) functions. By adjusting the parameters of these loss functions, we were able to reweight the hard and easy class examples. Each setup was evaluated on 4 folds, with uniformly distributed teeth with lesions. The highest lesion detection rate of 0.97 was achieved using FL, whereas FTL and CL achieved the detection rates of 0.92 and 0.88, respectively. The highest specificity value of 0.94 was achieved using CL, while the values of 0.88 and 0.89 were achieved using FL and FTL, respectively. The highest Dice score of 0.70 was achieved using the FTL and CL functions. The setup trained with FL achieved a slightly smaller value of 0.66 for the Dice score. By observing the predictions for individual lesions, we noticed a larger number of false negative predicted voxels and consequently lower Dice scores in images with very small or very large lesions. Small differences between predicted and ground truth segmentation maps have a significant impact on the Dice score for images with very small lesions, while a significant predicted portion of a lesion can result in a low Dice score for images with very large lesions. Since false negative predictions are less tolerable than false positive predictions in the lesion detection task, we conclude that the best performance of our method was achieved using the FL function with a sensitivity value of 0.97 and a specificity value of 0.88. In future work, we plan to explore the use of stronger anatomical constraints via generative models to improve the crucial teeth localization step.

When comparing our method with other SOTA methods, our approach achieved the highest sensitivity value and the same specificity value as \cite{Setzer20}. However, it is important to note that a direct comparison is not feasible due to the use of different datasets in the evaluation of these methods. 
Furthermore, all SOTA methods, except \cite{Orhan20}, utilized 2D slices rather than 3D volumes during the training and testing procedures. Additionally, the studies conducted in \cite{Setzer20} and \cite{Zheng20} were limited to small respective datasets. 
When considering the negative class, which refers to teeth without periapical lesions, only \cite{Setzer20} reported a specificity value. However, their dataset is highly selective as it consists of only 20 CBCT volumes with limited field-of-view. Moreover, among these volumes, there are 29 roots with lesions and 32 roots without lesions, thus creating a balanced distribution of roots with and without lesions, which does not reflect clinical practice.
%On the other hand, our method addresses the class imbalances commonly observed in real clinical data, as the majority of teeth in an average patient are healthy, and periapical lesions only occupy a small portion of CBCT images. Moreover, our method incorporates 3D volumes throughout all stages of the training and testing procedures, ensuring no loss of valuable information.

\begin{comment}
In contrast to SOTA methods for automatic periapical lesion detection in CBCT images, our method incorporates 3D volumes throughout all stages, ensuring no loss of valuable information. 
Additionally, to the best of our knowledge, we are the first to address the class imbalance issue associated with automatic periapical lesion detection. The class imbalance is commonly observed in clinical data, as periapical lesions only occupy a small portion of CBCT images. Therefore, it should be taken into consideration. 
Despite the challenges posed by CBCT images and periapical lesions, our fully automated two-step method achieved promising results in localizing teeth and detecting periapical lesions in CBCT data.
\end{comment}

\section{\uppercase{Conclusion}}
In this paper, we have presented a fully automated two-step method for detecting periapical lesions in CBCT images. In the first step, we utilize the 3D SpatialConfiguration-Net (SCN) for teeth localization. By using the teeth coordinates generated by the SCN, we extract relevant subregions from the original images, and use them to train the 3D U-Net for lesion segmentation in the second step. In contrast to other SOTA methods, our method incorporates 3D volumes throughout all stages, ensuring no loss of valuable information. Additionally, to the best of our knowledge, we are the first to address the class imbalance issue associated with automatic periapical lesion detection, which is commonly observed in clinical data. Despite the challenges posed by dental CBCT images, our method achieved promising results in localizing teeth and detecting periapical lesions in CBCT data.

\bibliographystyle{apalike}
{\small
\bibliography{example}}

%\section*{\uppercase{Appendix}}

%If any, the appendix should appear directly after the
%references without numbering, and not on a new page. To do so please use the following %command:
%\textit{$\backslash$section*\{APPENDIX\}}

\end{document}